# Synthesis of Borophene Nanoribbons on Ag(110) Surface

Qing Zhong[1,3], Longjuan Kong[1,3], Jian Gou[1,3], Wenbin Li[1,3], Shaoxiang Sheng[1,3], Shuo Yang[1,3], Peng Cheng[1,3], Hui Li[1,2]*, Kehui Wu[1,3,4]* and Lan Chen[1,3]*

[1]*Institute of Physics, Chinese Academy of Sciences, Beijing 100190, China*

[2]*Beijing Advanced Innovation Center for Soft Matter Science and Engineering, Beijing University of Chemical Technology, Beijing 100029, China*

[3]*School of physics, University of Chinese Academy of Sciences, Beijing 100049, China*

[4]*Collaborative Innovation Center of Quantum Matter, Beijing 100871, China*

*Email: hli@buct.edu.cn (H. L), khwu@iphy.ac.cn (K. W), lchen@iphy.ac.cn (L. C.)

**Abstract:** We present the successful synthesis of single-atom-thick borophene nanoribbons (BNRs) by self-assembly of boron on Ag(110) surface. The scanning tunneling microscopy (STM) studies reveal high quality BNRs: all the ribbons are along the [-110] direction of Ag(110), and can run across the steps on the surface. The width of ribbons is distributed in a narrow range around 10.3±0.2 nm. High resolution STM images revealed four ordered surface structures in BNRs. Combined with DFT calculations, we found that all the four structures of boron nanoribbons consist of the boron chains with different width, separated by hexagonal hole arrays. The successful synthesis of BNRs enriches the low dimensional allotrope of boron and may promote further applications of borophene.

**KEYWORD:** borophene, nanoribbon, Ag(110), scanning tunneling microscopy, molecular beam epitaxy

Since the discovery of graphene [1-3], the search for two-dimensional (2D) allotropes of other elements has attracted great interest. For examples, the 2D allotropes of other group IV elements, named as silicene [4-6], germanene [7-10] and stanene [11-13], have been predicted to be stable and realized by molecular beam epitaxy (MBE) growth in ultrahigh vacuum. Very recently, the 2D allotropes of boron (named as borophene), another neighbor of carbon in the periodic table elements, was also successfully synthesized on Ag(111) surface by two groups in parallel [14,15]. The experimentally observed borophene were found to consist of triangular boron grid and hexagonal holes, corresponding to $\beta_{12}$ sheet and $\chi_3$ sheet described by theoretical calculations [16-19], respectively. The fact that only a few structures were found on Ag(111) surface is surprising since freestanding borophene were predicted to be polymorphic with tremendous boron structures sharing competing binding energy near the global minimum [18]. It may indicate that the specific interfacial interactions between the substrate and borophene play important roles in determining the morphology of borophene. Thus, a more extensive investigation on the substrate effect on the boron structure is desired.

Patterning a 2D sheet into nanoribbons is another popular strategy to bring the new physical properties. For example, a size-dependent band gap can be induced by patterning a graphene sheet into nanoribbons [20,21]. The graphene community has spent great efforts to finally realize a controllable and clean fabrication of graphene nanoribbons [22-26]. Like graphene, borophene is also metallic, and borophene nanoribbons (BNRs) have been expected to host promising electronic properties [27-30]. But the experimentally synthesis of BNRs has still not been explored so far.

In our previous study on the growth of borophene on Ag(111) surface, we found that the borophene islands have predominately triangular shape and the edges are along the three crystallographic orientations of Ag(111). It is obvious that the six-fold symmetry of the Ag(111) surface results in the triangular shape of boron islands [15]. This stimulated our interest to explore the growth of boron sheet on an anisotropic substrate. Here we report a systematic study on the growth of borophene on the Ag(110) surface. We obtain two striking results: firstly we observed self-assembly of borophene sheet into uniform nanoribbons (BNRs), running along the [-110] direction of Ag(110). Secondly, four periodic structures were found on the BNR surfaces. STM combined with density functional theory (DFT) calculations revealed that all the four phases share similar, linear basic motifs with various widths, corresponding to the β and χ structures. We proposed a more accurate notation of these borophene structures, BC(n, m), where n and m denote to the number of atoms in widest and narrowest regions of a single boron chain. The high quality growth of BNRs on Ag(110) surface may open new possibilities in manipulation and application of low dimensional boron allotropes.

The experiments were performed in a home-built low-temperature STM combined with a MBE chamber with a base pressure of $2\times10^{-11}$ torr. Single crystal Ag(110) substrate was cleaned by repeated $Ar^+$ ion sputtering and annealing cycles. Boron atoms with 99.9999% purity was evaporated from an electron-beam evaporator onto the Ag(110) surface, while the surface was kept at a temperature of 570 K. After growth, the sample was transferred to STM chamber for characterization without breaking the vacuum. All the STM images were taken at 78 K, and the bias voltage was defined as the tip bias with respect to the sample.

Fig.1(a) is the STM image of a large area (100 × 100 nm$^2$) Ag(110) surface after boron growth. We observe several parallel nanoribbons, with width about ten nm, running throughout the image area along the [-110] direction of Ag(110). Larger area scans show that the borophene nanoribbons can run hundreds of nanometers. It is notable that the BNRs commonly cross steps of Ag(110) surface without losing the continuity, indicating that this monolayer sheet is itself a rigid structure. Fig.1(b) is the statistics of the width of BNRs obtained from STM images taken on different areas. Gaussian fitting nicely reproduces a width distribution of 10.3±0.2 nm. The narrow width distribution suggest that Ag(110) substrate is an ideal template for the formation of uniform BNRs. Nevertheless, we find that some BNR can split into two smaller ones, shown in Fig.1(c). It might be that the ribbon growth was obstructed by impurities during growth.

The atomic structures of BNRs can be revealed by high-resolution STM images. In Fig.1(c), the periodic patterns are different for two BNRs, and sometimes the patterns in different parts of a single BNR (for example, the upper part and lower part of right BNR) are also different. We have found four types of patterns with different unit cells in the BNRs, which are named as P1, P2, P3 and P4, respectively, for convenience. Fig.2(a-d) are the typical high resolution STM images of P1, P2, P3 and P4, respectively. The unit cell of P1 shown in Fig.2(a) is rectangular, with a lattice constants of $a$ = 0.40 nm and $b$ = 0.43 nm, where $b$ is along the

[-110] direction of Ag(110), as marked by the black arrow. Fig. 2(b) and 2(c) show that the unit cells of P2 and P3 of BNRs are rhombic, the angle between two base vectors being 61° for P2, and 60° for P3, respectively. Meanwhile, the lattice constants of P2 and P3 are $a$ = 0.44 nm, $b$ = 0.82 nm and $a$ = 0.45 nm, $b$ = 0.76 nm respectively. It is obvious that the unit cells of P2 and P3 are very similar. Considering the longer base vector of unit cells of both P2 and P3 are along the [-110] direction of Ag(110), P2 and P3 are image symmetry to each other. The unit cell of P4 shown in Fig.2(d) is also rectangular, but the lattice constants is $a$ = 0.89 nm and $b$ = 0.77 nm, which is about twice of those of P1. The unit cell of P4 had an angle of 30° with respect to the [-110] direction of Ag(110).

In order to determine the atomic structures of BNRs, we investigated a large number of 2D borophene structures from previous theoretical models [18] by overlapping them on Ag(110) with different rotation angles, and then compared the periodicities of the superlattices with experimental values. As a result, we found four types of striped structures which can nicely reproduce the unit cells observed in the STM images. To confirm the atomic structures of BNRs, we carried out first principles calculations on these four phases on Ag(110), and the optimized atomic models are shown in Fig. 3(a-h). Density functional theory (DFT) calculations were performed using projector-augmented wave (PAW) pseudopotentials in conjunction with the Perdew-Burke-Ernzerhof (PBE) [31] exchange function and plane-wave basis set with an energy cutoff at 330 eV. The computational model contains a boron monolayer on a four-layer Ag(110) substrate. We have kept the bottom two Ag layers fixed, and allow the topmost layers to relax until maximum forces were smaller than -0.01eV/ Å.

The vacuum region of more than 13 Å in the Z direction was applied, which is enough to eliminate the artificial period interaction. The two-dimentional Brillouin Zone was sampled using the Monkhorst-Pack scheme [32]. A 3×1 k-mesh was used for a 2×5 supercell (P1), while the density of k-meshes keeps for other supercells. All the calculations are performed with Vienna *ab initio* simulation package (VASP) [33]. After the relaxation of these structures, the simulated STM images are also calculated and shown in Fig.3(i-l), respectively.

The P1 phase can be perfectly explained by $\chi_3$ structure with the orientations of chains along the direction perpendicular to [-110] direction of Ag(110), as shown in Fig. 3(e). The borophene sheet is completely flat on Ag(110), demonstrating a high stability. This $\chi_3$ structure has been also been found on Ag(111) substrate [15], indicating it is an universal structure among all the 2D boron allotropes, which should result from the perfect commensuration between $\chi_3$ structure and Ag surface. The unit cell of P1 displayed in STM image (Fig. 2(a)) reflects both the distance between boron chains (4.33 Å) and distance between silver chains (4.08 Å) on Ag(110) surface, which is nice reproduced in our simulated STM images (Fig. 3(i)).

Based on the experimental STM images and the DFT calculations, P2 and P3 phases can both be explained by the β structure in Ref.18, but with mirror-symmetric orientations with respect to Ag(110), as shown in Fig 3(b) and (c). The optimized angles between boron ribbons and [-110] direction of Ag(110) is ±60.5º. Similar to P1 phase, the observed unit cells of P2 and P3 phases are also contributed by both inter-chain distances of boron ribbons and [-110] direction of Ag(110). The corresponding lattice constants ($a$ = 0.46 nm, $b$ = 0.77 nm) are consistent to the experimental measurements. In contrast to the perfectly planar configuration in P1, the β boron sheet is slightly buckled on Ag(110) due to the interfacial interaction, as displayed in Fig. 3(f) and (g). Such buckled structure can also well explain why the brightness of protrusions in STM images for P2 and P3 are not identical. The similar feature is also reproduced in simulated STM images (Fig. 3(j) and (k)), which further support our structure models. In addition, the experimental observed periodicities of BNRs are modulated by atomic chains of Ag(110) surface, which may indicate the high delocalization of electron density of 2D boron sheets.

The most possible atomic model for P4 is $β_8$ sheet predicted in Ref. 18, as shown in Fig. 3(d). The relaxed structure of P4 on Ag(110) exhibits larger fluctuation than P2 and P3, as shown in Fig. 3(h). Due to such large degree of buckling, the unit cell of P4 observed in experimental STM images ($a$ = 0.87 nm, $b$ = 0.77 nm) cannot reflect the periodicity of the substrate, but is solely attributed to the signature distances in $β_8$ boron sheet. As shown in Fig. 3(i), the simulated STM image is also in good agreement in experimental ones (Fig. 2(d)).

The calculated binding energies show that boron sheet in P1 (0.046 eV/Å$^2$) has stronger interaction with substrate as compared with the slightly bulked P2 (0.036 eV/Å$^2$) and P3 (0.036 eV/Å$^2$). Furthermore, the highly bulked P4 has the weakest interfacial interaction (0.035 eV/Å$^2$). The binding energy between boron sheets and Ag(110) is slightly larger than it on Ag(111) surface (~0.03 eV/Å$^2$) [15], which may originate from higher corrugation of Ag(110) than Ag(111).

Previously, we have reported two types of borophene structures on Ag(111), namely the $\chi_3$ and $\beta_{12}$ sheets [15,34]. Now, we have added up to this list another two phases, $\beta$ and $\beta_8$. Surprisingly, all these four kinds of structures share common features: they consist of full-filled boron chains separated by single rows of hexagonal holes. Considering that the Greek alphabet cannot give a direct sense of the structures of BNRs, we proposed here a more accurate description, Boron-Chain(n, m) [BC(n, m)], where n and m denote to the number of atoms in the widest and narrowest regions of a single boron chain. According to this classification, $\chi_3$, $\beta_{12}$, $\beta$ and $\beta_8$ sheet can be described as BC(2,2), BC(2,3), BC(3,4) and BC(4,4), respectively.

In summary, we successfully grew high quality boron nanoribbons on Ag(110) surface. These BNRs consist of boron chains named as BC(2,2), BC(3,4), and BC(4,4) with different orientation angles with respect to Ag(110) substrates. Considering that borophene on Ag(111) surface adopts BC(2,2) and BC(2,3) configurations, it is surprising that all borophene allotropes on Ag surface adopt very similar striped structures. Furthermore, the self-assembly of BNRs also indicates the strong template effect of the substrate on the growth of boron. The understanding of growth mechanism is desirable in order to obtain the borophene with particular structures as well as to explore the possible applications of borophene.

**Acknowledgement**

This work was supported by the MOST of China (grants nos. 2016YFA0300904, 2016YFA0202301, 2013CBA01601, 2013CB921702), the NSF of China (grants nos. 11674366, 11674368, 11334011, 11304368), and the Strategic Priority Research Program of the Chinese Academy of Sciences (grant no. XDB07020100).

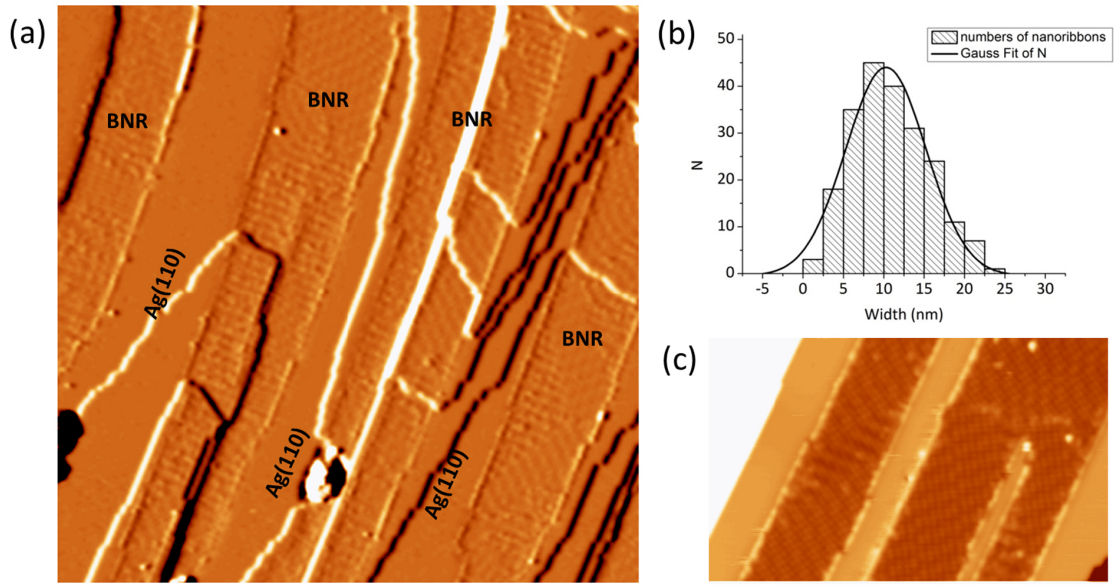

**Figure 1. Boron nanoribbons on Ag(110) surface.** (a) Derivative STM image shows boron nanoribbons grown on Ag(110). The image size is 100 × 100 nm$^2$. The nanoribbons run across the substrate steps without losing continuity. (b) A systematic statistics of nanoribbon width. Gaussian fitting is shown in a black line. (c) High-resolution STM image of two boron nanoribbons. The right one was split into two at lower right of the image. Atomic structure of boron nanoribbons was also revealed in this picture. Image size: 50 × 30 nm$^2$.

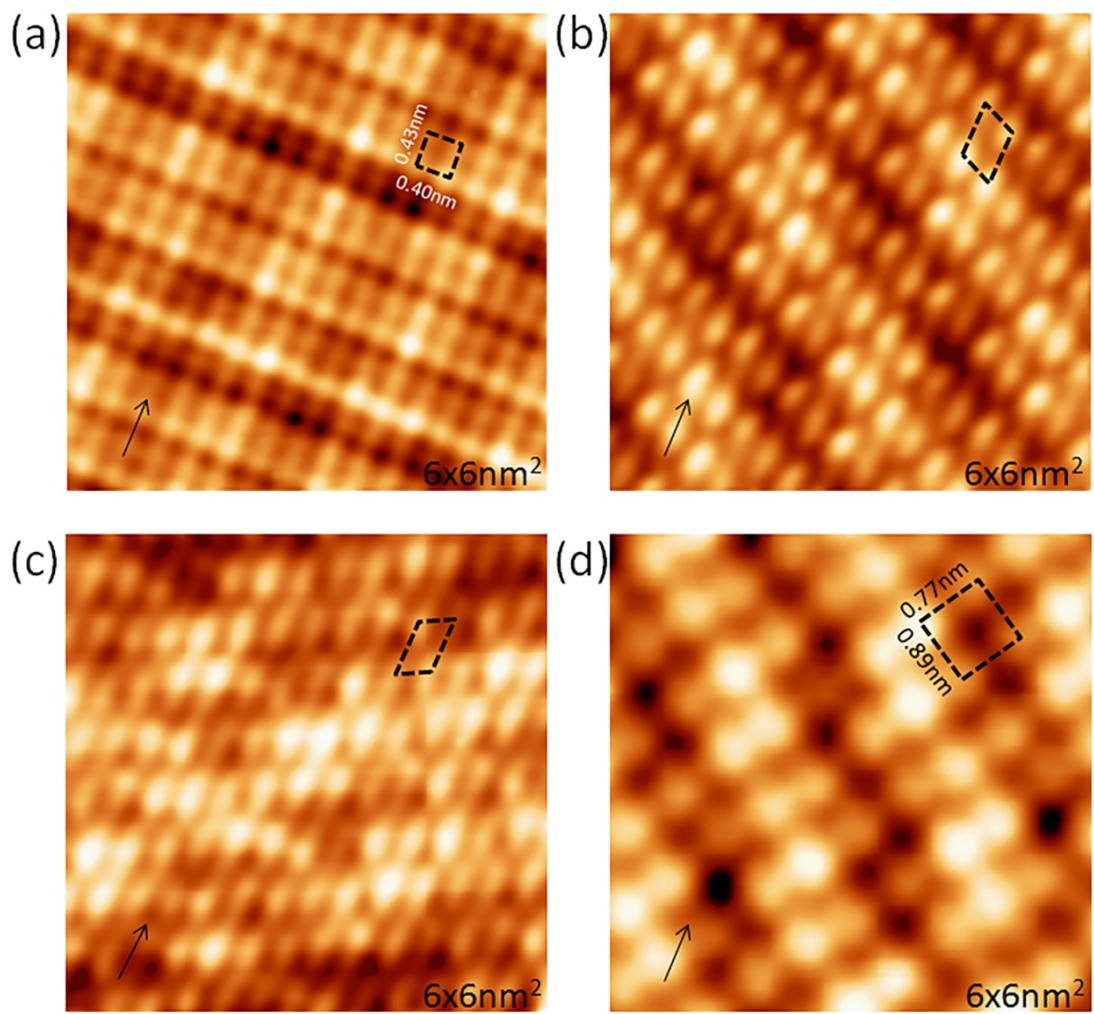

**Figure 2. High-resolution STM images of boron nanoribbons on Ag(110). (**a-d) are the high-resolution STM images for P1-P4, respectively. The unit cells of P1-P4 are marked by different dotted patterns. The size of the unit cells can be listed as follows. $a$ = 0.40 nm, $b$ = 0.43 nm, $\varphi$ = 90° for P1; $a$ = 0.44 nm, $b$ = 0.82 nm, $\varphi$ = 61° for P2; $a$ = 0.45 nm, $b$ = 0.76 nm, $\varphi$ = 60° for P3; $a$ = 0.89 nm, $b$ = 0.77 nm, $\varphi$ = 90° for P4. The direction of [-110] of Ag(110) are marked by black arrows in the images. The size of these images were 6 × 6 nm$^2$.

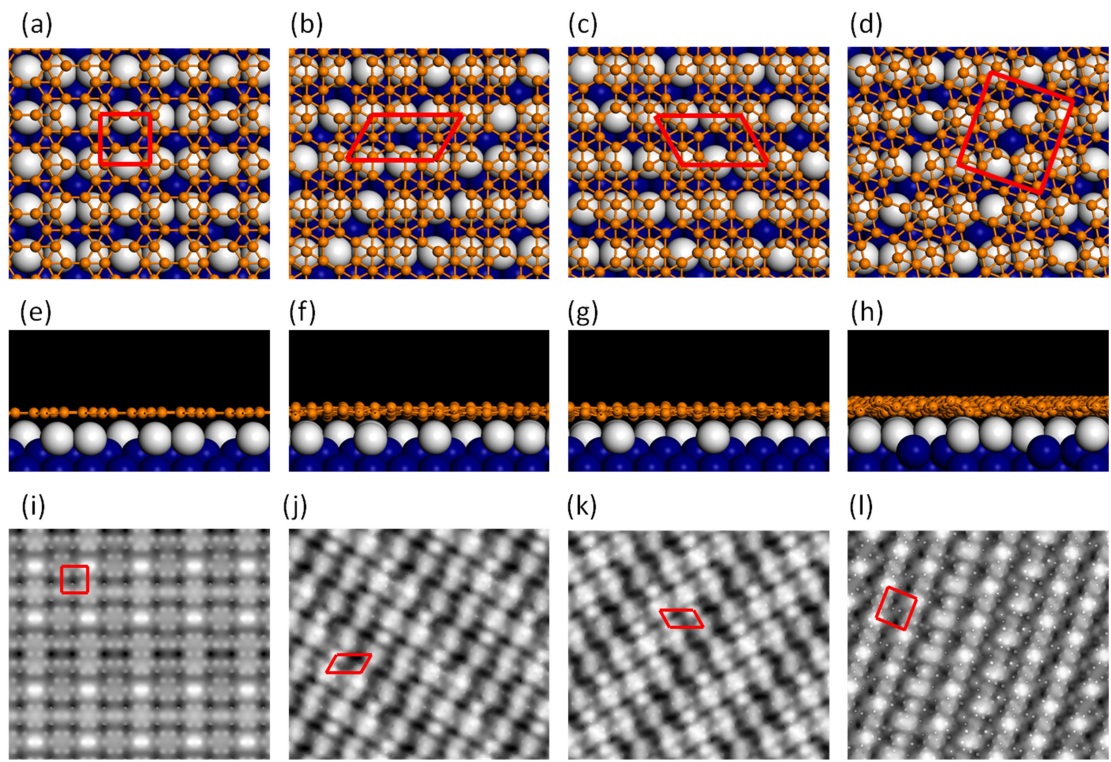

**Figure 3. Atomic structures of boron nanoribbons on Ag(110) optimized by DFT.** (a-d) Top and (e-h) side views of optimized P1-P4 BNRs on Ag(110) surface, respectively. Color codes: B, small orange spheres; Topmost Ag, large white spheres; Lower Ag, large blue spheres. (i-l) Simulated STM images for P1-P4 based on the calculated electronic density (0-2 eV respecting to the Fermi level). The red frameworks correspond to the observed unit cells in STM images.